# Who pumps spin current into nonmagnetic-metal (NM) layer in YIG/NM multilayers at ferromagnetic resonance?

Yun Kang[1], Hai Zhong[1], Runrun Hao[1], Shujun Hu[1], Shishou Kang[1]★, Guolei Liu[1], Y. Zhang[2], X. R. Wang[2]★, Shishen Yan[1], Yong Wu[3], Shuyun Yu[1], Guangbing Han[1], Yong Jiang[3] and Liangmo Mei[1]

[1]School of physics and State Key Laboratory of Crystal Materials, Shandong University, Jinan, Shandong, 250100, China.

[2]Physics Department, The Hong Kong University of Science and Technology, Clear Water Bay, Kowloon, Hong Kong, China.

[3]State Key Laboratory for Advanced Metals and Materials, School of Materials Science and Engineering, University of Science and Technology Beijing, Beijing 100083, China.

★email: skang@sdu.edu.cn; phxwan@ust.hk

**Spin pumping in Yttrium-iron-garnet (YIG)/nonmagnetic-metal (NM) layer systems under ferromagnetic resonance (FMR) conditions is a popular method of generating spin current in the NM layer. A good understanding of the spin current source is essential in extracting spin Hall angle of the NM and in potential spintronics applications. It is widely believed that spin current is pumped from precessing YIG magnetization into NM layer. Here, by combining microwave absorption and DC-voltage measurements on YIG/Pt and YIG/NM1/NM2 (NM1=Cu or Al, NM2=Pt or Ta), we unambiguously showed that spin current in NM came from the magnetized NM surface (in contact with YIG) due to the magnetic proximity effect (MPE), rather than the precessing YIG magnetization. This conclusion is reached through our unique detecting method where the FMR microwave absorption of the magnetized NM surface, hardly observed in the conventional FMR experiments, was greatly amplified when the electrical detection circuit was switched on.**

Spin current generation, detection, and manipulation involve fundamental science as well as the key-technologies[1,2] in spintronics. Spin pumping from precessing magnetization under ferromagnetic resonance (FMR) conditions is an attractive method for generating coherent pure spin current[3-8]. Pure spin current is the key resource in spintronics as well as a base for studying the inverse spin Hall effect (ISHE) characterized by the spin Hall angle $\alpha_{SH}$. Ferromagnetic-insulator (FI)/nonmagnetic-metal (NM) bilayers are believed to be the ideal settings for measuring $\alpha_{SH}$ of the non-magnetic metals. One of the widely studied such systems is Yttrium-iron-garnet (YIG)/Pt bilayer[9-16] because YIG [Y3Fe2(FeO4)3] is a well-known insulating magnetic material and Pt is a well-studied heavy metal with a strong spin-orbit interaction. The conventional understanding of the system under FMR conditions is that magnetization of YIG precesses coherently, and the precessing magnetization pumps pure spin current into Pt layer[9-12] across the YIG/Pt interface. The spin current in the Pt layer is then converted into a transverse (normal to spin flow direction and spin polarization) charge current that can be detected as an electrical voltage signal. YIG/Pt bilayer is believed to be a clean system for studying spin pumping and ISHE since YIG cannot conduct the electric current so that whatever the DC-voltage measured in Pt must be from the ISHE. The signature of spin pumping is the broadening of the FMR peak width of YIG as well as a detected DC-voltage in the Pt layer. Controversially, the extracted values of $\alpha_{SH}$ from different groups differ by two orders of magnitude, inconsistent with each other[10,16,17] in this "clean" system. It arises the question of who is responsible for the spin pumping, YIG or other magnetic sources? The answer to the question requires a good understanding of the interfacial phenomena between the YIG and a NM.

Here, we use YIG(16nm)/Pt(10nm) bilayer and YIG(16nm)/Cu(5nm)/Pt(10nm), YIG(16nm)/Cu(5nm)/Ta(10nm), YIG(16 nm)/Al(5nm)/Pt(10nm) trilayer systems in a rectangular cavity to investigate these issues by a combined measurements of microwave absorption and DC-voltage. Contrary to the popular belief, the spin current in Pt or Ta was not pumped from the precessing YIG magnetization, but from the magnetized NM surface (in contact with YIG) originated from the magnetic proximity effect (MPE)[18-24]. Interestingly and surprisingly, the FMR signal from the magnetized NM surface was greatly amplified when the electrical measurement circuit was connected (otherwise, the signal could hardly be observed). When the MPE was absent, such as in YIG(16nm)/Al(5nm)/Pt(10nm) samples, no DC-voltage signal was observed in Pt. Our experiments showed unambiguously that spin pumping from the insulating YIG layer into the metallic Pt or Ta layer *was not* efficient and effective, in comparison with that from a magnetized metallic surface into Pt or Ta.

## Results

**YIG(16nm)/Pt(10nm) systems.** The black circles in Fig. 1a are the derivative microwave absorption spectrum (*dI/dH*) of a pure YIG sample as a function of external magnetic field $H$. The lineshape of the FMR derivative absorption spectrum of the pure YIG sample follows a standard differential Lorentzian line with FMR peak at $H=2.497$ kOe, and peak width of $\Gamma=10$ Oe which shows high quality of our YIG samples[16].

The green/blue circles in Fig. 1a are the FMR derivative absorption spectra of a YIG(16nm)/Pt(10nm) bilayer strip when the electrical detection circuit is switched on/off (see the left inset of Fig. 1a and the methods below). In contrast, the FMR derivative microwave absorption spectrum of YIG/Pt bilayer sample appears to shift to a lower field with a seemingly broaden peak width that was observed in previous studies[5,16,25] and was used as an essential evidence of spin pumping from YIG. More strikingly, the microwave absorption signals are substantially different when the electrical detection circuit was switched on (green circles) and off (blue circles). Obviously, the absorption signal was greatly amplified when the electrical detection circuit was switched on. A more careful examination showed that the absorption curves of switch-off circuit are better described by two independent FMR signals.

One of them with a relative amplitude of $A_1$=60.5% was from the free YIG because it has the same peak position and peak width of $H_1$=2.497 kOe and $\Gamma_1$=10 Oe as the free YIG. The other with peak position $H_2$=2.488 kOe, peak width $\Gamma_2$=12 Oe, and relative amplitude $A_2$=39.5% was naturally attributed to the YIG covered by Pt. Because Pt modifies magnetic properties of YIG[25], the peak position and peak width of the second FMR signal differ slightly from those of the free YIG.

It is worthy to note that the FMR absorption curves of switch-on circuit were best fitted by three independent FMR signals as shown in Fig. 1b. Among the three signals in Fig. 1b, two signals (with $A_1$=54.5% and $A_2$=36.4%) are exactly those of free YIG and Pt-covered YIG, and the third signal of $H_3$=2.477 kOe, $\Gamma_3$=14 Oe, and $A_3$=9.1% came from the amplification of a very weak signal (hidden in the blue circles) originated from the MPE-induced magnetized Pt surface that was in contact with YIG. Furthermore, the corresponding DC-voltage detected in Pt was from the magnetized Pt surface since their peak positions and peak widths match exactly with each other as shown in Fig 1b.

To substantiate this interpretation, we fabricated also YIG/Pt bilayer samples in which YIG was fully covered by Pt layer. As shown in Fig. 1c, the blue and green circles are the FMR absorption signals when the electrical measurement circuit was switched on (green) and off (blue). As expected, the signal from the free YIG was absent. In Fig. 1d, the FMR absorption curves of switch-on circuit now consist of two signals respectively from the Pt-covered YIG and the magnetized Pt surface. Again, the DC-voltage relates to the signal of the magnetized Pt surface since their peak positions and widths match exactly with each other, and cannot be from the spin pumping of YIG.

The above conclusion could also be reached from the change of the shape of voltage-H curves as angle θ between ac-magnetic field and sample long edge varies. The DC-voltage originated from the spin pumping of YIG should follow a Lorentzian lineshape since the FMR absorption is described by the Lorentzian function[26]. Thus DC-voltage lineshape would be symmetric about its peak for any angle $\theta$ if the spin pumping was from YIG. However, as plotted in Fig. 2a, it is clear that most spectra consist of a superposition of a Lorentz- and a dispersive-type resonance lineshape[26,27] For a given $\theta$, the DC-voltage curve was fitted to Eq. (1) so that both symmetric and asymmetric components of the DC-voltages $U_{sym}$ and $U_{asy}$ were obtained. Their angle-dependences were plotted in Fig. 2b that fit well with theoretical prediction of Eq. (2) (see the Methods below). Voltages $U_{SR}^s$ (for symmetric component) and $U_{SR}^a$ (for asymmetric component) due to the spin rectification are $U_{SR}^s = 0.065\ \mu V$ and $U_{SR}^a = 0.568\ \mu V$. Voltage $U_{sp}$ from the ISHE due to spin pumping is $U_{sp} = 1.02\ \mu V$. Thus it shows that substantial amount of the DC-voltage came from the AMR and AHE of a ferromagnetic metal that resulted in an asymmetric lineshape, and the only possibility is that the Pt surface in contact with YIG was magnetized and generated a spin rectification voltage[28]. Obviously, the spin pumping effect generated a larger DC-voltage than the spin rectification. These results further confirmed the MPE and spin precession of magnetized Pt surface in YIG/Pt system.

Figure 3 is the $H$-dependence of the derivative microwave absorption with the switch-on circuit and DC voltage of a typical YIG/Pt strip sample for various frequencies at $\theta=0^o$. The DC voltages (Fig. 3a) have a symmetric Lorentzian shape while the microwave absorption has multi-peaks (Fig. 3b) due to different FMR sources. This implies that only one FMR source pumped spin current into Pt and generated the DC-voltage. Clearly, the peak position and width of the DC-voltage match well with those of the tiny FMR signal. Furthermore, Fig. 3c shows that the peak field of the DC-voltage increased with the frequency that fits well with the Kittel formula[29,30], $f=(\gamma/2\pi)(H_{res}*(H_{res}+4\pi M_s))^{1/2}$, with the gyromagnetic ratio $\gamma = g\mu_B/h=1.738\times10^{11}$ T$^{-1}$ s$^{-1}$ and the saturation magnetization $M_s =0.248$ T. It gave a Lander factor $g=2.08$ for magnetized Pt surface. The fitted $M_s$ of magnetized Pt is very closed to the observed value in Ni/Pt system by XMCD measurement[19]. Thus, this result further supports our assertion that the precessing magnetization of YIG did not pump spin current to Pt layer, contrary to the popular belief[5,14-16,31].

**YIG(16nm)/Cu(5nm)/Pt(10nm) and YIG(16nm)/Cu(5nm)/Ta(10nm) systems.**
To further substantiate our claim that the DC-voltage in YIG/Pt bilayer is due to the spin pumping from the MPE-induced magnetized Pt layer, a 5nm thick Cu was inserted between YIG and Pt (Ta) so that Pt (Ta) surfaces were not in contact with YIG and no MPE is possible for Pt (Ta). The upper panel of Fig. 4 is the typical FMR derivative microwave absorption spectrum of one of our YIG/Cu/Pt samples. The blue circles are the results when the electrical detection circuit was switched off while the green circles are those when the circuit was switched on. The signal from the magnetized Pt surface was obviously absent, and was replaced by a new FMR signal at an even lower field of $H=2.46$ kOe, far below YIG resonance peak, and with a peak width of $\Gamma=12$ Oe. Although this new signal was seen in a 50 times enlarged figure as shown by the black circles in the top panel, it was extremely weak when the electrical detection circuit was switched off. Similar to YIG/Pt bilayer samples, an extra signal can be clearly observed when the electrical detection circuit was switched on. As expected this time, the DC voltage of YIG/Cu/Pt and/or YIG/Cu/Ta (middle and lower panels of Fig. 4) was observed at *H=2.46* kOe, exactly corresponding to the new FMR signal. The peak widths of the FMR and DC-voltage signals matched again with each other as shown in the middle (for YIG/Cu/Pt) and lower (for YIG/Cu/Ta) panels. In contrast, there were no DC-voltage signals at the YIG resonant fields, confirming that the DC-voltage was not due to the spin pumping of YIG, but due to that of the magnetized Cu surface (in contact with YIG) [32,33,34,35]. The signs of the DC-voltages of YIG/Cu/Pt and YIG/Cu/Ta samples are opposite due to the opposite sign of spin Hall angle for Pt (positive) and Ta (negative)[16].

Again, above experiments do not support the general belief that precessing YIG magnetization at FMR pumps spin current into Pt or Ta in YIG/Pt, YIG/Cu/Pt, and YIG/Cu/Ta systems. Our results are consistent with the assertion that it was the magnetized NM surface pumping spin current into the Pt or Ta layer. The clear evidences include 1) DC-voltage peaks were far from the FMR peaks of the YIG, but were exactly overlapped with the FMR peak of MPE-induced magnetized NM surface; 2) the angle-dependence of DC-voltage lineshape that shows big contribution from AMR and AHE of a magnetized metal. Furthermore, the MPE of both Pt and Cu were confirmed by our first-principle calculations (see the Methods). It was found that a few Cu atomic layers adjacent to Ni was magnetized with average moment about *-0.02 $\mu_B$/atom*, which was only about 1/5 of the average moment of Pt (0.11 $\mu_B$/atom) in Ni(111)/Pt system[19]. If one assumes that Cu/Ni and Cu/YIG (Pt/Ni and Pt/YIG) have the similar MPE, then the precession of this small moment can pump spin current into Pt or Ta layer. We can naturally interpret our observed DC-voltage as the ISHE. This smaller magnetized Cu moment explains the much smaller voltage than that in YIG/Pt system as shown in Figs. 1 and 2. The small negative Cu magnetic moment is also consistent with lower resonance field for the magnetized Cu due to the negative exchange field at FMR[29].

**YIG(16nm)/Al(5nm)/Pt(10nm) systems.** To further verify the assertion that spin current in NM layer(s) was not from YIG in YIG/NM1 bilayer or YIG/NM1/NM2 multilayer samples, but from magnetized NM1 surface (in contact with YIG), we did a controlled experiment with YIG(16nm)/Al(5nm)/Pt(10nm) samples. Al has no MPE, in consistent with our first-principle calculations on Ni/Al system. As shown in Fig. 5, the derivative microwave absorption spectrum of a typical YIG/Al/Pt samples does not change whether the electrical detection circuit was switched on or off, in contrast to that for YIG/Pt or YIG/Cu/Pt(Ta) systems. Consistent with our assertion, there was no spin current in Pt layer since YIG could not pump detectable spin current and no DC-voltage was observed as shown in the bottom panel of Fig. 5.

## Discussion

A magnetic metallic film at its FMR can generate not only a DC signal but also a radio-frequency ac field[28,36] by the AHE and the AMR. As illustrated in Fig. 6 (see the Methods) when the sample is connected to Cu connection-pads through two Al wires, the whole structure becomes a patch antenna and a high frequency pass filter. According to the patch antenna theory[37,38], the ac signals from the AMR and AHE of the magnetic metallic layer and from ISHE of nonmagnetic metallic layer can be radiated through fringing fields at the radiating edges. This will result in the amplification of the FMR signal of the metallic film. Here we termed this amplification as "antenna effect" when the Al wires were connected to Cu-Pads (see experimental setup, switch-on circuit). Fig. 6d shows clearly this antenna effect since the microwave absorption of a 3nm-thick Permalloy (Py) film sample was greatly enhanced when Al wires were connected to Cu-Pads.

One possible reason, that the precessing YIG could not inject a detectable spin current into Pt layer, might be due to the mismatch in the electronic structures of YIG and Pt, resulting in an inefficient angular momentum transfer. The FMR linewidth broadening and additional damping mechanism observed previously may be due to the overlap of resonant peaks of both YIG and magnetized Pt[12,14]. Thus one should extract the spin Hall angle $\alpha_{SH}$ by taking into the account of the new findings reported here.

## Conclusion

In summary, our experiments on YIG/Pt bilayer and YIG/Cu/Pt (YIG/Cu/Ta) trilayer samples showed that the FMR microwave absorption was mainly from three sources: free YIG, YIG covered by a NM, and the magnetized NM surface arising from the MPE. Interestingly, the FMR microwave absorption signal from the magnetized NM layer was pronounced only when the electrical detection circuit was switched on. The electrical detection circuit acted as an antenna for the FMR signal of the magnetized NM surface. Surprisingly, the DC-voltages were from the spin rectification effects and spin pumping of the magnetized NM layers, instead of spin pumping of YIG alone. Thus, contrary to the popular belief, our studies suggest that precessing magnetization of YIG does not pump detectable spin current into the NM layer. Our findings are very important for properly extracting the spin Hall angle and for a better understanding of the concept of interface mixing conductance[9-11,16].

## Methods

**Sample preparation and experimental procedure.** YIG [$Y_3Fe_2(FeO_4)_3$] films (16 nm) were fabricated on $Gd_3Ga_5O_{12}$ (GGG) wafers by pulsed laser deposition (PLD). X-ray diffraction (XRD) and atomic force microscopy (AFM) showed that our YIG are high quality (See supplementary Figure 1). Py, Pt, Cu, Ta or Al with high purity (4N) was then deposited on YIG by magnetron sputtering to create a NM strip with a mask of *0.2* mm × 2.3 mm. All samples were cut into 1 mm × *3* mm for DC-voltage and microwave absorption measurements in a homemade X-band microwave absorption spectrometer.

The experimental setup is shown in Fig. 6. FMR microwave absorption and DC-voltage were measured at frequency *f=9.7* GHz of TE10 mode in the X-band cavity. The sample with size of *1* mm x *3* mm was mounted in the middle of a shorted copper plate at one end of the cavity that can rotate in the XY-plane as illustrated in Fig. 6. The angle between the Y-axis and the long edge of the sample is denoted as θ. Two thin rectangular copper sheets of *1.5* mm × *8* mm were symmetrically placed on the both sides (*1* mm away from sample) of sample in the cavity as illustrated in Fig. 6. These two small copper sheets were isolated from the shorted copper plate and acted as electrical connection pads that connected to a SR-530 lock-in amplifier of Stanford Research Systems or Keithley-2182 nanovoltmeter. It should be pointed out that these two thin copper sheets did not affect the X-band microwave distribution from the angular dependence measurements of FMR for Py (see Supplementary Fig.2). Two Al wires of diameter about 30 micrometers were attached to two long-edge ends of the sample. As illustrated in Fig. 6, the electrical measurement circuit was switched on when the other ends of Al wires were connected to the two Cu-pads. The in-plane external field and ac microwave field were always orthogonal with each other in order to have a maximal precessing magnetization.

**Angular dependence:** The DC-voltage from ISHE, AMR and/or AHE of a ferromagnetic metal near the FMR has a symmetric Lorentz-lineshape and an asymmetric dispersive-lineshape [26,27,28]

$$U = U_{sym} L(H, H_0, \Gamma) + U_{asy} D(H, H_0, \Gamma)$$

$$L(H, H_0, \Gamma) = \frac{\Gamma^2}{(H-H_0)^2 + \Gamma^2} \quad (1)$$

$$D(H, H_0, \Gamma) = \frac{\Gamma(H-H_0)}{(H-H_0)^2 + \Gamma^2}$$

Here, $H_0$ and $\Gamma$ are respectively the resonance field and resonant peak width. $U_{sym}$ and $U_{asy}$ are the voltages of the symmetry and asymmetry components of DC-voltage that depend on angle $\theta$ as[39]

$$U_{sym} = U_{SR}^s \sin(2\theta)\sin(\theta) + U_{SP} \cos(\theta) \quad (2)$$

$$U_{asy} = U_{SR}^a \sin(2\theta)\sin(\theta)$$

Here, $U_{SR}^s$, $U_{SR}^a$ are the voltages due to the spin rectification. $U_{sp}$ is the voltage from the ISHE due to spin pumping.

**First-Principle calculations:** Because our computation resources do not allow us to perform reliable calculations on YIG/Pt and/or YIG/Cu systems due to the huge unit cells, we performed the calculations on Ni(111)/Cu systems instead by using the same method in Reference 19 for calculating MPE of Pt in Ni(111)/Pt systems.

# Acknowledgements

This work was supported by the National Basic Research Program of China under Grant No. 2015CB921502 and 2013CB922303, the National Natural Science Foundation of China under Grant Nos 11474184, 11174183 and 11504203, and the 111 project under Grant No. B13029. YZ and XRW were supported by the Hong Kong RGC Grants No. 163011151 and No. 605413. YW and YJ were supported by the National Natural Science Foundation of China under Grant No. 51501007.

# Author contributions

X.R.W. and S.K. contributed to project design and manuscript writing; Y.K., H.Z. and R.H. carried out the experiments; S.H. performed the first principle calculations; Y.W.

Figure legends

Fig. 1 (a) The FMR derivative absorption spectra of free YIG and YIG/Pt bilayer strip with $\theta=0^o$ ($\theta$ is the angle between microwave field and the sample long edges). The lower panel is the corresponding DC-voltage (bottom) spectra of YIG/Pt bilayer schematically illustrated on the right. The left inset is the schematic diagram of the electrical. The right inset shows the sample structure YIG/Pt stripe. (b) The fit of FMR spectrum of YIG/Pt strip obtained with antenna effect by three FMR signals respectively for free YIG, YIG covered with Pt, and magnetized Pt surface in contact with YIG. The DC signal agrees with the assertion of spin pumping from the magnetized Pt surface. (c) The FMR derivative microwave absorption (upper) and DC-voltage (lower) spectra of fully covered YIG/Pt bilayer (illustrated in the lower left) with $\theta=170^o$. The inset shows the YIG sample fully covered by Pt. (d) The FMR spectrum of fully covered YIG/Pt bilayer with the antenna effect is best fitted by two FMR signals from YIG covered with Pt and magnetized Pt surface. The corresponding DC-voltage signal was from the magnetized Pt surface.

Fig. 2 (a) The DC-voltage spectra of a YIG(16nm)/Pt (10nm) strip at various angle $\theta$. The symmetrical (b) and asymmetrical (c) components of DC-voltages defined in Eq. (1) were extracted. The solid lines are the best fits of Eq. (2) with $U_{SR}^s = 0.065 \ \mu V$, $U_{SR}^a = 0.568 \ \mu V$ and $U_{sp} = 1.02 \ \mu V$. The inset illustrates experimental setup and angle $\theta$.

Fig. 3 The H-dependence of DC-voltage (a) and FMR spectra (b) of YIG/Pt stripe line at $\theta=0^o$ and for various frequencies with antenna effect. (c) The frequency dependence of peak position of DC-voltages. The solid line is the fit to the Kittel formula.

Fig. 4 The FMR spectra of a YIG/Cu bilayer sample (upper panel) and DC-voltage spectra of YIG/Cu/Pt (middle panel) and YIG/Cu/Ta (lower panel) with $\theta=0^o$. In the upper panel, the green (blue) circles are the FMR derivative microwave absorption spectra when the electrical detection circuit is switched on (off). A weak signal, as shown by the black circles that is the zoom-in (50 times enlarged) of the blue circles inside the red rectangle, was amplified when the electrical circuit was switched on. The DC-voltage in Pt (middle panel) and Ta (lower panel) can be fitted well by the Lorentzian function (solid lines).

Fig. 5 Top: The FMR derivative microwave absorption spectra of a YIG/Al(5nm)/Pt(10nm) sample when the electrical detection circuit was switched on (red) and off (blue). No MPE-induced magnetized Al was observed. Bottom: No DC-voltage signal in Pt was observed.

Fig. 6 (a) The experimental setup for the microwave absorption measurement and electrical detection of FMR in which DC-voltage along the long edge of the sample was measured. (b) The zoom-in of sample and copper-pads. The total thickness of the sample including GGG substrate is about 1 mm. (c) The in-plane rotation geometry of sample. The microwave of frequency $f=9.7$GHz propagated along the Z-axis, and the external field H was along the X-axis. $e$ and $h$ are the electric and magnetic components of the microwave, respectively along the X-axis and the Y-axis. The angle between the Y-axis and the long edge of sample was denoted as $\theta$ that varies as the shorted copper plate rotates in the XY-plane. (d) FMR signals of 3 nm Py thin film strip when the Al wires were connected/disconnected to the Cu-pads (with/without antenna effect). The inset is the equivalent circuit with antenna effect.

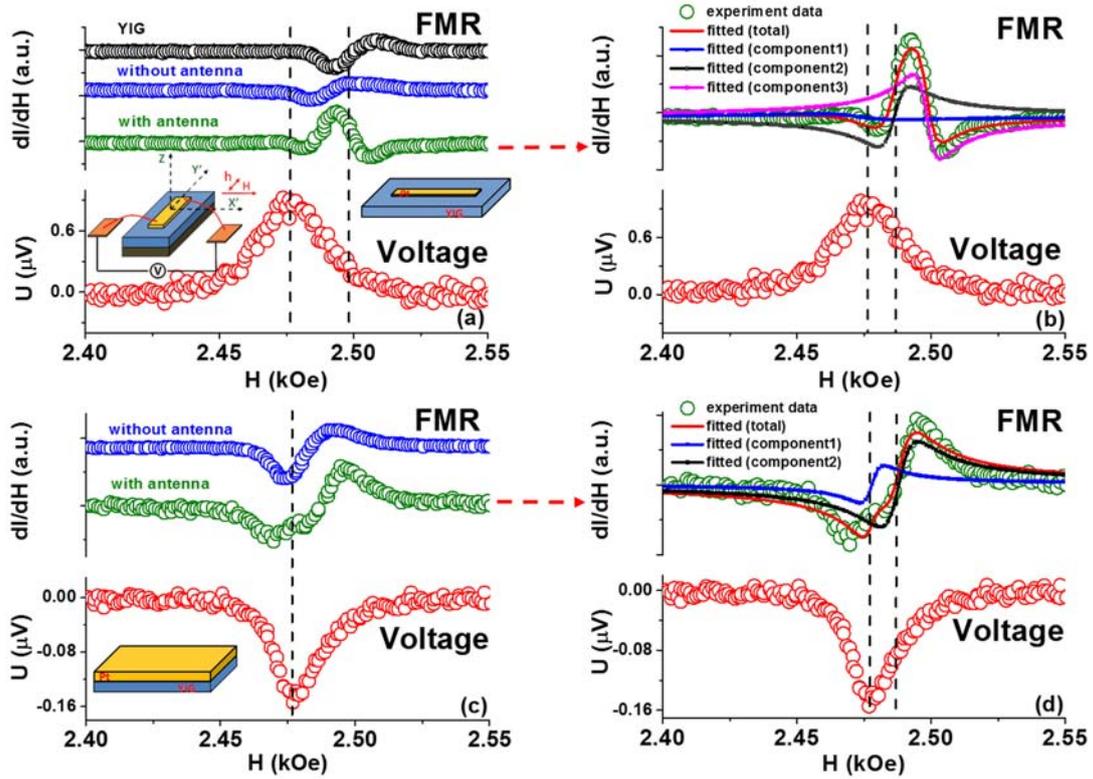

Fig. 1 Kang *et al*.

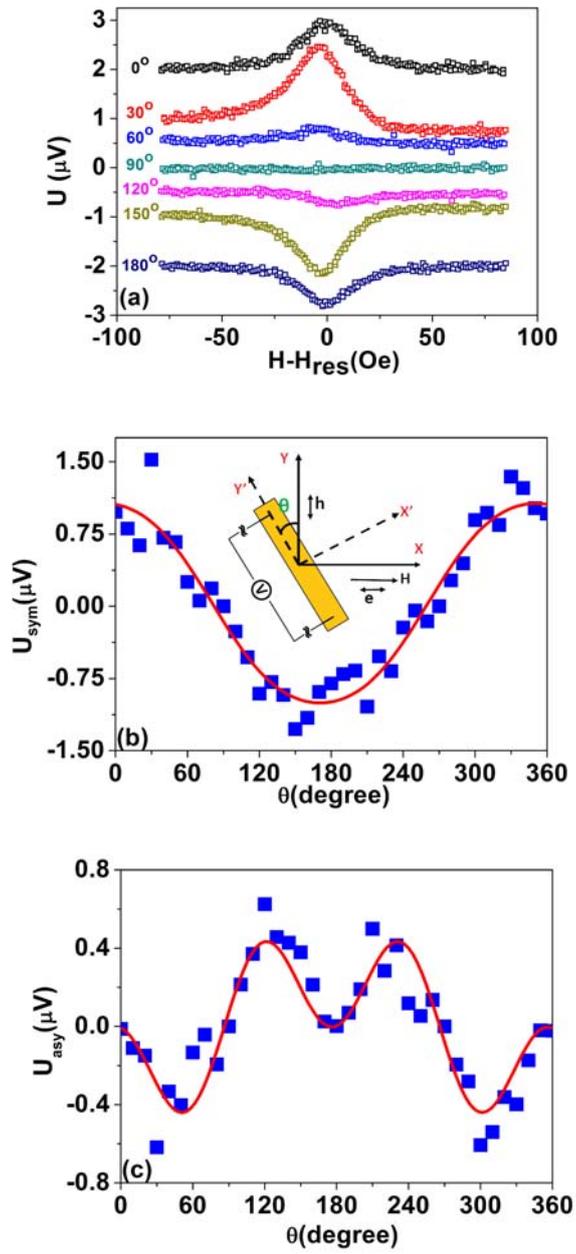

FIG. 2 Kang *et al*.

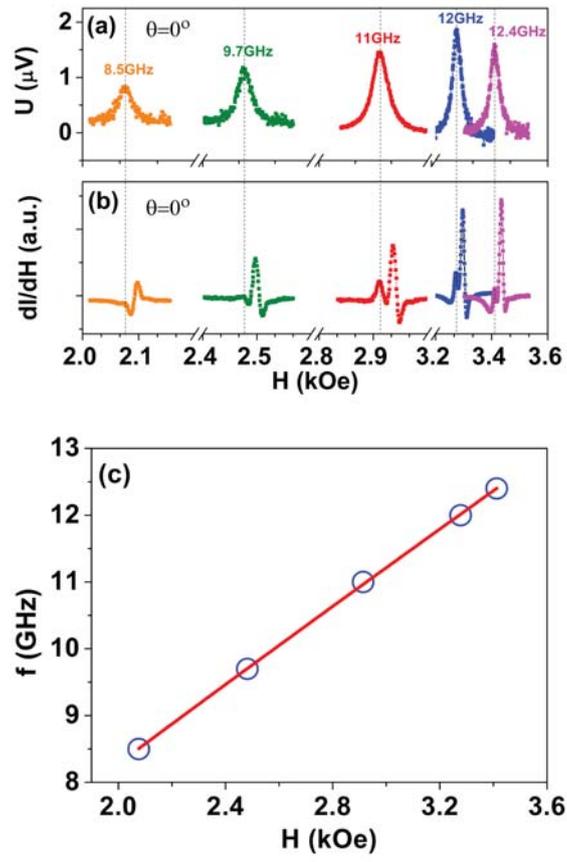

FIG. 3 Kang *et al*.

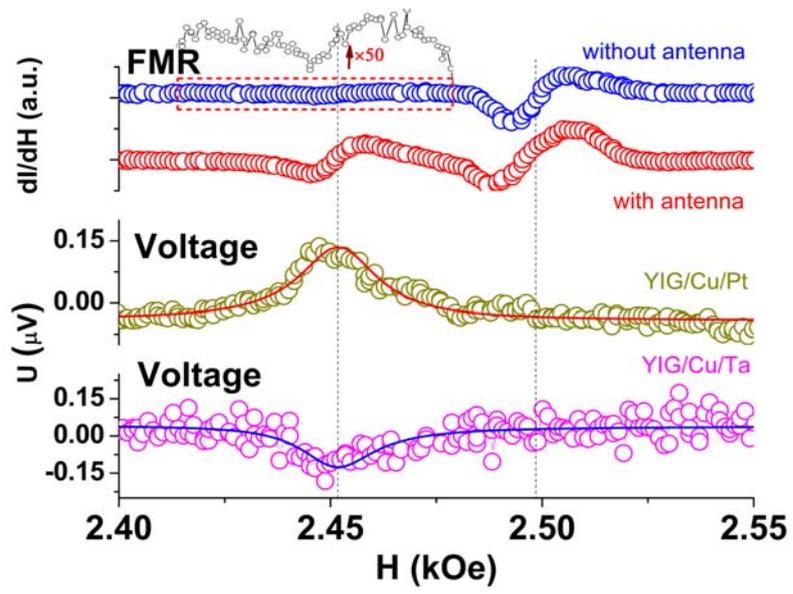

FIG. 4 Kang *et al*.

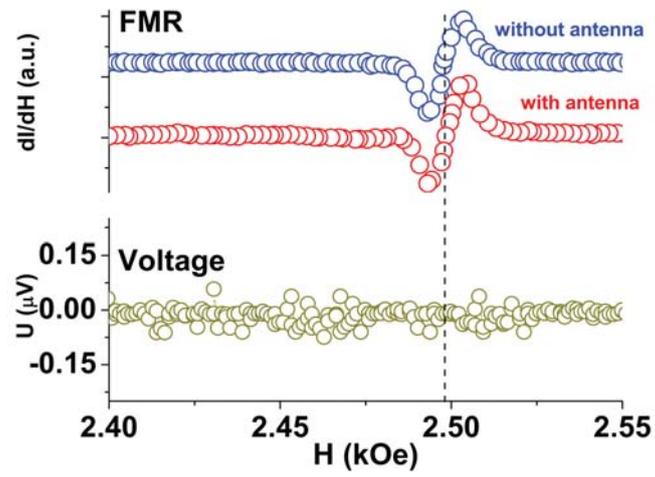

FIG. 5 Kang *et al*.

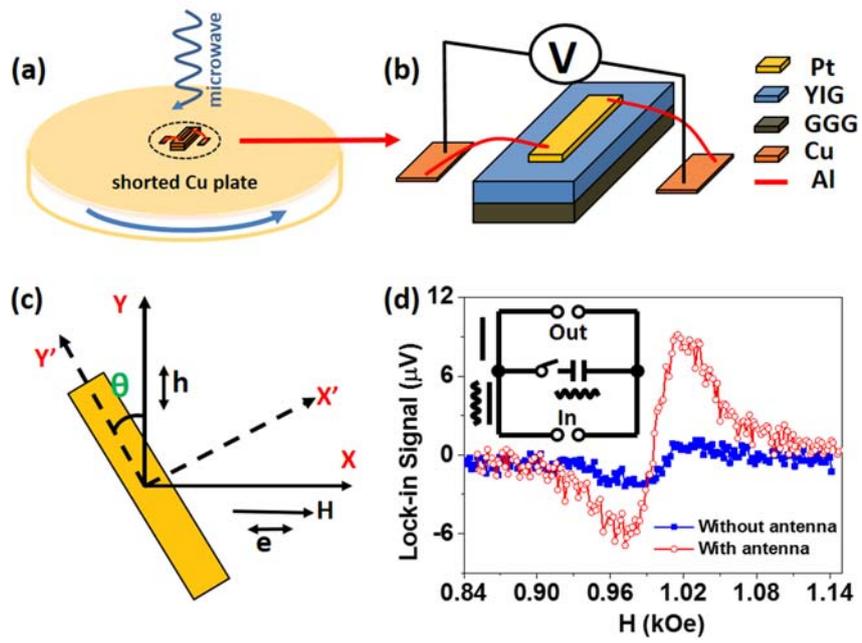

FIG 6 Kang *et al*.